\documentclass[12pt]{article}

\usepackage{mathrsfs}
\usepackage[T1]{fontenc}
\usepackage{mathpazo}
\usepackage{setspace}
\usepackage{amsfonts}
\usepackage{amssymb}
\usepackage{amsmath}
\usepackage{bm}
\usepackage{esint}                   
\usepackage{epsfig}
\usepackage{latexsym}
\usepackage{color}
\usepackage{graphicx}
\usepackage{cite}
\usepackage{hyperref}
\usepackage{nicefrac}
\usepackage[latin1]{inputenc}
\usepackage{pstricks}
\usepackage{slashed}
\usepackage{multirow}
\usepackage{pifont}
\usepackage{caption}

\usepackage{tikz}
\usetikzlibrary{arrows,matrix,positioning,shapes,arrows}
\usetikzlibrary{shapes.geometric, arrows, calc, intersections}

\def\at{\tilde{\a}}
\def\bt{\tilde{\b}}
\def\gt{\tilde{\g}}

\def\Dt{\tilde{\D}}
\def\fe{\mathfrak{e}}
\def\he{\hat{e}}

\def\hybrid{\topmargin -30pt    \oddsidemargin 0pt 
        \headheight 0pt \headsep 0pt
        \textwidth 6.25in       
        \textheight 9.5in       
        \marginparwidth .875in
        \parskip 5pt plus 1pt   \jot = 1.5ex}

\hybrid

\def\baselinestretch{1.2}

\catcode`\@=11

\def\marginnote#1{}
\newcount\hour
\newcount\minute
\newtoks\amorpm
\hour=\time\divide\hour by60
\minute=\time{\multiply\hour by60 \global\advance\minute by-\hour}
\edef\standardtime{{\ifnum\hour<12 \global\amorpm={am}%
        \else\global\amorpm={pm}\advance\hour by-12 \fi
        \ifnum\hour=0 \hour=12 \fi
        \number\hour:\ifnum\minute<10 0\fi\number\minute\the\amorpm}}
\edef\militarytime{\number\hour:\ifnum\minute<10 0\fi\number\minute}

\def\draftlabel#1{{\@bsphack\if@filesw {\let\thepage\relax
   \xdef\@gtempa{\write\@auxout{\string
      \newlabel{#1}{{\@currentlabel}{\thepage}}}}}\@gtempa
   \if@nobreak \ifvmode\nobreak\fi\fi\fi\@esphack}
        \gdef\@eqnlabel{#1}}
\def\@eqnlabel{}
\def\@vacuum{}
\def\draftmarginnote#1{\marginpar{\raggedright\scriptsize\tt#1}}

\def\draft{\oddsidemargin -.5truein
        \def\@oddfoot{\sl preliminary draft \hfil
        \rm\thepage\hfil\sl\today\quad\militarytime}
        \let\@evenfoot\@oddfoot \overfullrule 3pt
        \let\label=\draftlabel
        \let\marginnote=\draftmarginnote
   \def\@eqnnum{(\theequation)\rlap{\kern\marginparsep\tt\@eqnlabel}%
\global\let\@eqnlabel\@vacuum}  }

\def\draft2{
        \def\@oddfoot{\sl preliminary draft \hfil
        \rm\thepage\hfil\sl\today\quad\militarytime}
        \let\@evenfoot\@oddfoot \overfullrule 3pt
        \let\label=\draftlabel
        \let\marginnote=\draftmarginnote
   \def\@eqnnum{(\theequation)\rlap{\kern\marginparsep\tt\@eqnlabel}%
\global\let\@eqnlabel\@vacuum}  }

\def\preprint{\twocolumn\sloppy\flushbottom\parindent 2em
        \leftmargini 2em\leftmarginv .5em\leftmarginvi .5em
        \oddsidemargin -.5in    \evensidemargin -.5in
        \columnsep .4in \footheight 0pt
        \textwidth 10.in        \topmargin  -.4in
        \headheight 12pt \topskip .4in
        \textheight 6.9in \footskip 0pt
        \def\@oddhead{\thepage\hfil\addtocounter{page}{1}\thepage}
        \let\@evenhead\@oddhead \def\@oddfoot{} \def\@evenfoot{} }

\def\numberbysection{\@addtoreset{equation}{section}
        \def\theequation{\thesection.\arabic{equation}}}

\def\underline#1{\relax\ifmmode\@@underline#1\else
        $\@@underline{\hbox{#1}}$\relax\fi}

\def\titlepage{\@restonecolfalse\if@twocolumn\@restonecoltrue\onecolumn
     \else \newpage \fi \thispagestyle{empty}\c@page\z@
        \def\thefootnote{\fnsymbol{footnote}} }

\def\endtitlepage{\if@restonecol\twocolumn \else \newpage \fi
        \def\thefootnote{\arabic{footnote}}
        \setcounter{footnote}{0}}  

\catcode`@=12
\relax

\def\figcap{\section*{Figure Captions\markboth
        {FIGURECAPTIONS}{FIGURECAPTIONS}}\list
        {Figure \arabic{enumi}:\hfill}{\settowidth\labelwidth{Figure
999:}
        \leftmargin\labelwidth
        \advance\leftmargin\labelsep\usecounter{enumi}}}
 \relax
\def\tablecap{\section*{Table Captions\markboth
        {TABLECAPTIONS}{TABLECAPTIONS}}\list
        {Table \arabic{enumi}:\hfill}{\settowidth\labelwidth{Table
999:}
        \leftmargin\labelwidth
        \advance\leftmargin\labelsep\usecounter{enumi}}}
 \relax
\def\reflist{\section*{References\markboth
        {REFLIST}{REFLIST}}\list
        {[\arabic{enumi}]\hfill}{\settowidth\labelwidth{[999]}
        \leftmargin\labelwidth
        \advance\leftmargin\labelsep\usecounter{enumi}}}
 \relax
\makeatletter
\newcounter{pubctr}
\def\publist{\@ifnextchar[{\@publist}{\@@publist}}
\def\@publist[#1]{\list
        {[\arabic{pubctr}]\hfill}{\settowidth\labelwidth{[999]}
        \leftmargin\labelwidth
        \advance\leftmargin\labelsep
        \@nmbrlisttrue\def\@listctr{pubctr}
        \setcounter{pubctr}{#1}\addtocounter{pubctr}{-1}}}
\def\@@publist{\list
        {[\arabic{pubctr}]\hfill}{\settowidth\labelwidth{[999]}
        \leftmargin\labelwidth
        \advance\leftmargin\labelsep
        \@nmbrlisttrue\def\@listctr{pubctr}}}
 \relax
\makeatother

\def\be{\begin{equation}}
\def\ee{\end{equation}}
\def\ba{\begin{eqnarray}}
\def\ea{\end{eqnarray}}

\def\r{\rho}
\def\a{\alpha}

\def\b{\beta}

\def\g{\gamma}
\def\G{\Gamma}
\def\d{\delta}
\def\D{\Delta}
\def\e{\epsilon}

\def\m{\mu}
\def\n{\nu}
\def\om{\omega}
\def\Om{\Omega}
\def\l{\lambda}

\def\s{\sigma}

\def\cH{{\cal H}}

\def\no{\noindent}

\def\IR{\relax{\rm I\kern-.18em R}}

\def\inv{^{\raise.0ex\hbox{${\scriptscriptstyle -}$}\kern-.05em 1}}

\begin{document}


\renewcommand{\theequation}{\thesection.\arabic{equation}}
\csname @addtoreset\endcsname{equation}{section}

\begin{titlepage}
\begin{center}

\hfill HU-EP-23/59

\phantom{xx}
\vskip 0.5in

{\large \bf Supersymmetric solutions of type-II supergravity from $\l$-deformations and zoom-in limits}

\vskip 0.5in

{\bf Georgios Itsios}

Institut f\"{u}r Physik, Humboldt-Universit\"{a}t zu Berlin,\\
IRIS Geb\"{a}ude, Zum Gro{\ss}en Windkanal 2, 12489 Berlin, Germany\\

\vskip .2in


\end{center}

\vskip .4in

\centerline{\bf Abstract}

\no We construct the embedding of the $\l$-model on $SL(2 , \mathbb{R}) \times SU(2) \times SU(2)$ in type-II supergravity. In the absence of deformation, the ten-dimensional background corresponds to the near-horizon limit of the NS1-NS5-NS5 brane intersection. We show that when the deformation is turned on, supersymmetry breaks by half, and the solution preserves 8 supercharges. The Penrose limits along two null geodesics of the deformed geometry are also considered. It turns out that none of the associated plane-wave backgrounds exhibit supernumerary supercharges.

\vfill
\no
 {
georgios.itsios@physik.hu-berlin.de
}

\end{titlepage}
\vfill
\eject



\def\baselinestretch{1.2}
\baselineskip 20 pt

\newcommand{\eqn}[1]{(\ref{#1})}


\tableofcontents

\section{Introduction}

Supergravity solutions on $AdS$ spaces often arise as the near-horizon limit of brane intersections. One such example comes from the NS1-NS5-NS5 brane system, whose geometry near the horizon is $AdS_3 \times S^3 \times S^3 \times S^1$ \cite{Cowdall:1998bu}. This constitutes a part of a type-II supergravity solution, which is also supported by a Neveu-Schwarz (NS) three-form with components proportional to the volume forms of $AdS_3$ and the two $S^3$'s. Such a configuration is half-maximally supersymmetric \cite{Cowdall:1998bu}.

Furthermore, the metric on $AdS_3 \times S^3 \times S^3$, along with the three-form flux, can be viewed as a WZW model \cite{Witten:1983ar} on $SL(2, \mathbb{R}) \times SU(2) \times SU(2)$. Deformations of WZW models on semi-simple groups that preserve integrability have garnered considerable attention in the literature. Examples of such deformations include the $\l$-model \cite{Sfetsos:2013wia} and generalisations of it \cite{Georgiou:2016zyo,Georgiou:2017jfi,Georgiou:2018hpd,Georgiou:2018gpe,Driezen:2019ykp,Hollowood:2014rla,Hollowood:2014qma,Sfetsos:2017sep}.

Integrable $\s$-models can be an important playground for studying new paradigms of the AdS/CFT correspondence \cite{Maldacena:1997re}. In this regard, it is useful to promote the $\s$-model background fields to a full supergravity solution. That means constructing the appropriate dilaton and Ramond-Ramond (RR) fields that are necessary for the equations of motion to be satisfied. In the context of the $\l$-model, this was first achieved in \cite{Sfetsos:2014cea} for the deformation based on $SL(2, \mathbb{R}) \times SU(2)$, as well as deformations based on symmetric spaces. Nevertheless, in the case of $SL(2, \mathbb{R}) \times SU(2)$, the outcome was a solution of the type-IIB* supergravity \cite{Hull:1998vg}, and as such, it is of no interest for our considerations.\footnote{More type-II supergravity solutions for $\l$-deformations based on (super) cosets have been constructed in \cite{Demulder:2015lva, Borsato:2016zcf, Itsios:2019izt, Chervonyi:2016ajp, Borsato:2016ose}. In addition, embeddings for the closely related $\eta$-deformed models \cite{Klimcik:2002zj,Klimcik:2008eq,Delduc:2013fga,Delduc:2013qra} have been found in \cite{Hoare:2015wia,Hoare:2015gda,Hoare:2018ngg,Lunin:2014tsa,Hoare:2022asa}.}
This problem for the embedding of the $SL(2 , \mathbb{R}) \times SU(2)$ $\l$-model has been successfully addressed recently in \cite{Itsios:2023kma} by an analytic continuation of the $SL(2, \mathbb{R})$ parameters. As a result, the first 1/4 supersymmetric type-II solution from a $\l$-deformed model has been found.

In the present work, we construct the $\l$-deformation for the near-horizon limit of the NS1-NS5-NS5 setup, and we show that it is 1/4 supersymmetric. The background presented in \cite{Itsios:2023kma} can be obtained by a zoom-in limit that makes one of the deformed three-spheres flat. We also consider the Penrose limits \cite{Penrose1976} around two null geodesics of the deformed geometry and demonstrate the non-existence of supernumerary supercharges for the pp-wave solutions.

The plan of the paper is as follows: In Section \ref{LambdaDeformedSolution}, we construct the 1/4 supersymmetric type-IIA solution and we lift it to eleven dimensions. We also consider various zoom-in limits and discuss the relationship with the $\l$-deformed $AdS_3 \times S^3 \times T^4$ background. Section \ref{SupersymmetryAnalysisSynopsis} is devoted to the supersymmetry analysis of the deformed type-IIA solution. In Section \ref{PenroseLimits}, we discuss two Penrose limits and the supersymmetry of the associated plane-wave solutions. Conclusions and future ideas are contained in Section \ref{ConclusionsAndFutureIdeas}. We have also included four appendices. In Appendix \ref{LambdaBackgroundFields}, we collect the background fields for the $\l$-deformed $\s$-models on $SL(2 , \mathbb{R})$ and $SU(2)$. Appendix \ref{UndeformedSolutions} offers a review of the undeformed pure NS and type-IIA/IIB pure RR backgrounds on $AdS_3 \times S^3 \times S^3 \times S^1$. Appendix \ref{TypeIIBsolution} contains the type-IIB counterpart of the $\l$-deformed solution discussed in Section \ref{LambdaDeformedSolution}. Details on the Killing spinor equations are collected in Appendix \ref{DetailsOnSupersymmetry}.

\section{The type-IIA deformed solution}
\label{LambdaDeformedSolution}

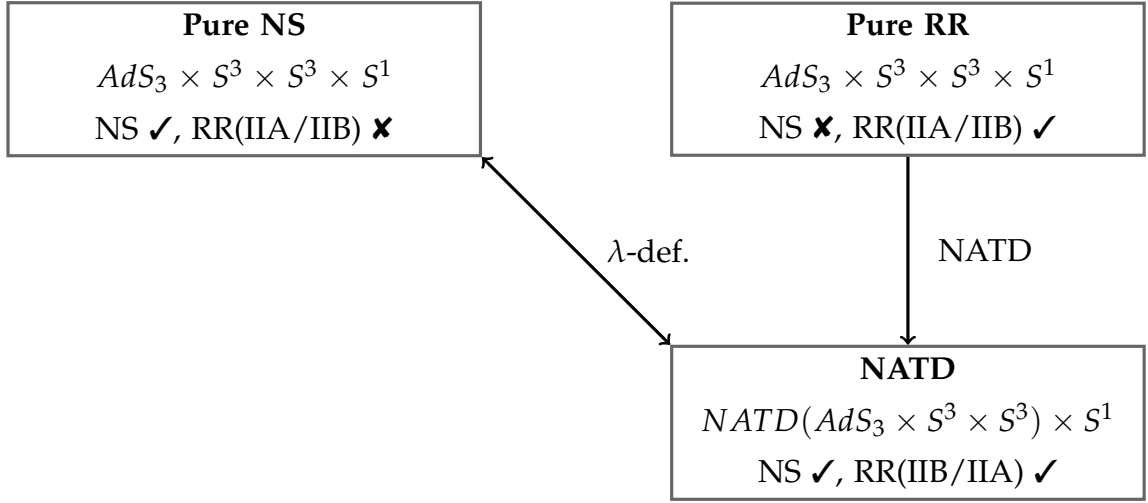
\begin{figure}[!h]
        \centering
        \begin{tikzpicture}
        [
        squarednode/.style={%
            rectangle,
            draw=black!60,
            fill=white,
            very thick,
            minimum size=5mm,
            text centered,
            text width=6cm,
        }
        ]
        \node[squarednode]      (NATD)                                                  {{\bf NATD} \\[5pt] $NATD(AdS_3 \times S^3 \times S^3) \times S^1$ \\[5pt] NS \ding{51}, RR(IIB/IIA) \ding{51}};
        \node[squarednode]      (pureRR)       [above=2.5cm of NATD]  {{\bf Pure RR} \\[5pt] $AdS_3 \times S^3 \times S^3 \times S^1$ \\[5pt] NS \ding{56}, RR(IIA/IIB) \ding{51}};
        \node[squarednode]      (pureNS)       [left=2.5cm of pureRR]    {{\bf Pure NS} \\[5pt] $AdS_3 \times S^3 \times S^3 \times S^1$ \\[5pt] NS \ding{51}, RR(IIA/IIB) \ding{56}};

        \draw[very thick, ->] (pureRR.south) -- node[anchor=west] {\,\, NATD} (NATD);
        \draw[very thick, <->] (pureNS.south east) -- node [right,midway] {\,\, $\l$-def.}(NATD.north west);
        \end{tikzpicture}
        \captionsetup{width=0.8\textwidth}
        \caption{Relation of the various solutions. Notice that NATD maps a type-IIA theory to a type-IIB one and vice versa. The pure NS solution corresponds to $\l = 0$. On the other hand, when $\l$ approaches the identity one finds the NATD backgrounds.}
        \label{figure:solutionsWeb}
    \end{figure}

In this section we describe how to embed in type-II supergravity two copies of the $\l$-deformed $\s$-model on $SU(2)$ together with that on $SL(2 , \mathbb{R})$. Our approach resembles the idea of \cite{Sfetsos:2014cea,Itsios:2023kma} for $AdS_3 \times S^3 \times T^4$. In particular, we want to construct a solution that interpolates between the pure NS background on $AdS_3 \times S^3 \times S^3 \times S^1$ and the non-Abelian T-dual (NATD) of its pure RR cousin (type-IIA or type-IIB). The non-Abelian T-duality takes place in both three-spheres as well as the $AdS_3$. For this reason, if the pure RR background is a type-IIA solution its NATD will be a type-IIB one and vice versa. The aforementioned backgrounds and their relations are illustrated in figure \ref{figure:solutionsWeb}. Below we give the details of the $\l$-deformed embedding while the pure NS and RR solutions are listed in Appendix \ref{UndeformedSolutions}.

\noindent\textbf{The metric.} Staring with the pure RR solution we derive the metric of the deformed geometry by three simple replacements. More precisely, we substitute the line element on $AdS_3$ by the one in \eqref{SL2RLambdaBackgroundFields} and we do the same for the two three-spheres, where instead we use two copies of the line element in \eqref{SU2LambdaBackgroundFields}. The resulting metric has now the form
\begin{equation}
 \label{DeformedMetric}
 \begin{aligned}
  ds^2 & = L^2_0 \, k \Big( \frac{1 + \l}{1 - \l} \, d\at^2 + \frac{1 - \l^2}{\tilde{\D}(\at)} \cosh^2\at \big( d\bt^2 - \cosh^2\bt \, d\gt^2 \big) \Big)
  \\[5pt] &
  + L^2_1 \, k \Big( \frac{1 + \l}{1 - \l} \, d\a^2_1 + \frac{1 - \l^2}{\D(\a_1)} \sin^2\a_1 \big( d\b^2_1 + \sin^2\b_1 \, d\g^2_1 \big) \Big)
  \\[5pt] &
  + L^2_2 \, k \Big( \frac{1 + \l}{1 - \l} \, d\a^2_2 + \frac{1 - \l^2}{\D(\a_2)} \sin^2\a_2 \big( d\b^2_2 + \sin^2\b_2 \, d\g^2_2 \big) \Big) + d\om^2 \, .
 \end{aligned}
\end{equation}
Here the functions $\D$ and $\tilde{\D}$ are defined in \eqref{Delta} and \eqref{DeltaTilde} respectively. Clearly when $\l = 0$, the metric \eqref{UndeformedMetric} is recovered, representing the geometry on $AdS_3 \times S^3 \times S^3 \times S^1$. The latter is recognised as the spacetime that arises from the near-horizon limit of the NS1-NS5-NS5 brane setup \cite{Cowdall:1998bu}. However, the presence of the deformation breaks the isometry of $AdS_3 \times S^3 \times S^3$. Nonetheless, a subspace with topology $AdS_2 \times S^2 \times S^2$ is still present.

For the better presentation of the RR fields later, we also introduce the orthogonal frame
\begin{align}
  & e^0 = L_0 \sqrt{k \frac{1 - \l^2}{\tilde{\D}(\at)}} \cosh\at \, \sinh\bt \, d\gt , \,\,
     e^1 = L_0 \sqrt{k \frac{1 - \l^2}{\tilde{\D}(\at)}} \cosh\at \, d\bt , \,\,
     e^2 = L_0 \sqrt{k \frac{1 + \l}{1 - \l}} d\at ,
  \nonumber\\[5pt]
  & e^3 = L_1 \sqrt{k \frac{1 + \l}{1 - \l}} d\a_1 , \,\,
     e^4 = L_1 \sqrt{k \frac{1 - \l^2}{\D(\a_1)}} \sin\a_1 \, d\b_1 , \,\,
     e^5 = L_1 \sqrt{k \frac{1 - \l^2}{\D(\a_1)}} \sin\a_1 \, \sin\b_1 \, d\g_1 ,
  \nonumber\\[5pt]
  & e^6 = L_2 \sqrt{k \frac{1 + \l}{1 - \l}} d\a_2 , \,\,
     e^7 = L_2 \sqrt{k \frac{1 - \l^2}{\D(\a_2)}} \sin\a_2 \, d\b_2 , \,\,
     e^8 = L_2 \sqrt{k \frac{1 - \l^2}{\D(\a_2)}} \sin\a_2 \, \sin\b_2 \, d\g_2 ,
  \nonumber\\[5pt]
  & e^9 = d\om \, .
 \label{DeformedFrame}
\end{align}

\noindent\textbf{The NS form.} Unlike the pure RR solution where the NS form vanishes, the deformed solution has a non-trivial two-form. More specifically, there are three contributions coming from the deformations of the $AdS_3$ and the two three-spheres

\begin{equation}
 \label{DeformedTwoForm}
 \begin{aligned}
  B_2 & = L^2_0 \, k \Big( \at + \frac{(1 - \l)^2}{\tilde{\D}(\at)} \cosh\at \, \sinh\at \Big) \cosh\bt \, d\bt \wedge d\gt
  \\[5pt] &
  + L^2_1 \, k \Big( - \a_1 + \frac{(1 - \l)^2}{\D(\a_1)} \cos\a_1 \, \sin\a_1 \Big) \sin\b_1 \, d\b_1 \wedge d\g_1
  \\[5pt] &
  + L^2_2 \, k \Big( - \a_2 + \frac{(1 - \l)^2}{\D(\a_2)} \cos\a_2 \, \sin\a_2 \Big) \sin\b_2 \, d\b_2 \wedge d\g_2 \, .
 \end{aligned}
\end{equation}
As it is expected, when $\l = 0$ the field strength of $B_2$ reduces to \eqref{UndeformedNSform}.

\noindent\textbf{The dilaton.} The deformed solution also supports a non-trivial dilaton
\begin{equation}
 \label{DeformedDilaton}
 \Phi = - \frac{1}{2} \ln \Big( \tilde{\D}(\at) \D(\a_1) \D(\a_2) \Big) \, .
\end{equation}
Obviously this vanishes if we set $\l = 0$.

\noindent\textbf{The RR forms.} A naive approach to obtain the RR fluxes is to write down an ansatz inspired from the NATD version of the pure RR solution and then solve the supergravity equations of motion. However, here we will follow a more systematic procedure proposed in \cite{Sfetsos:2014cea}. According to this, one can build the RR poly-form of the deformed solution, $\widehat{\mathbb{F}}$, from the poly-form of the pure RR background, $\mathbb{F}$, through the following relation
\begin{equation}
 e^{\Phi} \, \widehat{\slashed{\mathbb{F}}} = \m(\l) \, \slashed{\mathbb{F}} \, \Om^{-1} \, .
\end{equation}
The slash in the above expression denotes contraction with the $\G$-matrices. Moreover, $\Om$ is a matrix that can be written in terms of the $\G$'s and $\m$ is a constant that depends on the deformation parameter. In our case, the matrix $\Om$ (expressed in the basis \eqref{DeformedFrame}) reads
\begin{equation}
 \begin{aligned}
  \Om = \frac{1}{\sqrt{\tilde{\D}(\at) \D(\a_1) \D(\a_2)}} & \Big( (1 - \l) \sinh\at \G^{012} + (1 + \l) \cosh\at \G^2 \Big)
  \\[-5pt]
   \times & \Big( (1 - \l) \cos\a_1 \G^{345} - (1 + \l) \sin\a_1 \G^3 \Big)
  \\[5pt]
   \times & \Big( (1 - \l) \cos\a_2 \G^{678} - (1 + \l) \sin\a_2 \G^6 \Big) \, .
 \end{aligned}
\end{equation}
Also, the constant $\m$ has the form
\begin{equation}
 \m = \frac{2 \l}{\sqrt{k (1 - \l^2)} (1 + \l)} \, .
\end{equation}
Notice that for $\l = 0$ the constant $\m$ vanishes and therefore one finds the pure NS solution as it is anticipated. On the other hand, when $\l \ne 0$ the deformation generates RR fluxes. The form of the matrix $\Om$ implies that if we start with a type-IIA solution then the deformed fluxes are in type-IIB and vice versa. With this in mind, we construct fluxes both in type-IIA and in type-IIB for the deformed theory. It turns out that the two theories are related via T-duality in the $\om$ direction. For this reason, we provide the RR content of the type-IIA solution below, whereas we quote the details of the type-IIB one in Appendix \ref{TypeIIBsolution}.

For the type-IIA RR fields we find that the zero-form (Romans mass) vanishes while the rest are
\begin{equation}
 \label{DeformedRR2Form}
 \begin{aligned}
   F_2 & = 2 \m (1 - \l) (1 + \l)^2 \Big(\frac{1}{L_0} \sinh\at \sin\a_1 \sin\a_2 e^{36}
   - \frac{1}{L_1} \cosh\at \cos\a_1 \sin\a_2 e^{26}
   \\[5pt] &
   + \frac{1}{L_2} \cosh\at \sin\a_1 \cos\a_2 e^{23} \Big)
   - 2 \m (1 + \l) (1 - \l)^2 \Big(\frac{1}{L_0} \cosh\at \cos\a_1 \cos\a_2 e^{29}
   \\[5pt] &
   - \frac{1}{L_1} \sinh\at \sin\a_1 \cos\a_2 e^{39}
   - \frac{1}{L_2} \sinh\at \cos\a_1 \sin\a_2 e^{69} \Big)
 \end{aligned}
\end{equation}
and
%
%
 \begin{align}
  F_4 = & - 2 \m (1 + \l)^3 \cosh\at \sin\a_1 \sin\a_2 \Big( \frac{e^{0136}}{L_0} - \frac{e^{2456}}{L_1} + \frac{e^{2378}}{L_2} \Big)
  \nonumber\\[5pt] &
  + 2 \m (1 - \l)^3 \sinh\at \cos\a_1 \cos\a_2 \Big( \frac{e^{0129}}{L_0} + \frac{e^{3459}}{L_1} + \frac{e^{6789}}{L_2} \Big)
  \nonumber\\[5pt] &
  + 2 \m (1 - \l) (1 + \l)^2 \left( \cosh\at \sin\a_1 \cos\a_2 \Big( \frac{e^{2459}}{L_0} - \frac{e^{0139}}{L_1} \Big) \right.
  \label{DeformedRR4Form}
  \\[5pt] &
  \left. + \cosh\at \cos\a_1 \sin\a_2 \Big( \frac{e^{2789}}{L_0} - \frac{e^{0169}}{L_2} \Big)
  - \sinh\at \sin\a_1 \sin\a_2 \Big( \frac{e^{3789}}{L_1} + \frac{e^{4569}}{L_2} \Big) \right)
  \nonumber\\[5pt] &
  + 2 \m (1 + \l) (1 - \l)^2 \left( \sinh\at \cos\a_1 \sin\a_2 \Big( \frac{e^{3456}}{L_0} + \frac{e^{0126}}{L_1} \Big) \right.
  \nonumber\\[5pt] &
  \left. + \sinh\at \sin\a_1 \cos\a_2 \Big( \frac{e^{3678}}{L_0} - \frac{e^{0123}}{L_2} \Big)
  - \cosh\at \cos\a_1 \cos\a_2 \Big( \frac{e^{2678}}{L_1} - \frac{e^{2345}}{L_2} \Big) \right) \, .
  \nonumber
 \end{align}
For the sake of economy we adopt the notation $e^{a_1 \ldots a_p} = e^{a_1} \wedge \ldots \wedge e^{a_p}$. Notice that the above field content solves the equations of motion for the type-IIA supergravity under the condition \eqref{CondRadii}. Also, since the Romans mass vanishes this background can be lifted to eleven dimensions. Later we will give explicit formulas for the uplift.

\subsection{The $T^4$ limit}
\label{T4Limit}

A $\l$-deformed $AdS_3 \times S^3 \times T^4$ type-IIB solution that preserves $8$ supercharges was constructed recently in \cite{Itsios:2023kma}. Here we derive its type-IIA cousin by applying the limit \eqref{FromS3toT4} in the supergravity background of the previous section. The two solutions on the $\l$-deformed $AdS_3 \times S^3 \times T^4$ geometry are related
\footnote{
Up to a shift in the dilaton followed by an appropriate rescaling of the RR fields.
}
via a T-duality in one of the torus directions. Below we present this background in more detail.

\noindent\textbf{The metric.} Taking the limit \eqref{FromS3toT4} in \eqref{DeformedMetric} gives
\begin{equation}
 \label{DeformedT4Metric}
 \begin{aligned}
  ds^2 & = L^2_0 \, k \Big( \frac{1 + \l}{1 - \l} \, d\at^2 + \frac{1 - \l^2}{\tilde{\D}(\at)} \cosh^2\at \big( d\bt^2 - \cosh^2\bt \, d\gt^2 \big) \Big)
  \\[5pt] &
  + L^2_1 \, k \Big( \frac{1 + \l}{1 - \l} \, d\a^2_1 + \frac{1 - \l^2}{\D(\a_1)} \sin^2\a_1 \big( d\b^2_1 + \sin^2\b_1 \, d\g^2_1 \big) \Big)
  \\[5pt] &
  + k \, \frac{1 + \l}{1 - \l} \, \Big( d\r^2 + \r^2 \, \big( d\b^2_2 + \sin^2\b_2 \, d\g^2_2 \big) \Big) + d\om^2 \, .
 \end{aligned}
\end{equation}
The first three terms in the third line above parametrise a Euclidean three-dimensional space, $\mathbb{R}^3$, and can be replaced by a three-torus, $T^3$. Subsequently, the last can be combined with $d\om^2$ to give a $T^4$.

Notice that the frame components $e^6 , \, e^7$ and $e^8$ in \eqref{DeformedFrame} are mapped as follows
\begin{equation}
 \label{DeformedFrameT4}
 e^6 \mapsto \sqrt{k \frac{1 + \l}{1 - \l}} d\r , \qquad
 e^7 \mapsto \sqrt{k \frac{1 + \l}{1 - \l}} \r \, d\b_2 , \qquad
 e^8 \mapsto \sqrt{k \frac{1 + \l}{1 - \l}} \r \, \sin\b_2 \, d\g_2 \, .
\end{equation}
Meanwhile, the rest of the components are not affected by the limit.

\noindent\textbf{The dilaton.} In the limit \eqref{FromS3toT4} the function $\D(\a_2)$ becomes a $\l$-dependent constant. As a result the dilaton now is
\begin{equation}
 \label{DeformedT4Dilaton}
 \Phi = - \frac{1}{2} \ln \Big( \tilde{\D}(\at) \D(\a_1) (1 - \l)^2 \Big) \, .
\end{equation}
\noindent\textbf{The NS form.} After applying \eqref{FromS3toT4}, the third line in \eqref{DeformedTwoForm} vanishes. Therefore we are left with
\begin{equation}
 \label{DeformedTwoFormT4}
 \begin{aligned}
  B_2 & = L^2_0 \, k \Big( \at + \frac{(1 - \l)^2}{\tilde{\D}(\at)} \cosh\at \, \sinh\at \Big) \cosh\bt \, d\bt \wedge d\gt
  \\[5pt] &
  + L^2_1 \, k \Big( - \a_1 + \frac{(1 - \l)^2}{\D(\a_1)} \cos\a_1 \, \sin\a_1 \Big) \sin\b_1 \, d\b_1 \wedge d\g_1 \, .
 \end{aligned}
\end{equation}

\noindent\textbf{The RR forms.} It is easy to see that when taking \eqref{FromS3toT4}, the terms in \eqref{DeformedRR2Form} and \eqref{DeformedRR4Form} that contain the radius $L_2$ or the function $\sin\a_2$ vanish. Then for the RR two-form we obtain
\begin{equation}
 \label{DeformedRR2FormT4}
  F_2 = - 2 \m (1 + \l) (1 - \l)^2 \Big(\frac{1}{L_0} \cosh\at \cos\a_1 e^{29}
  - \frac{1}{L_1} \sinh\at \sin\a_1 e^{39} \Big) \, .
\end{equation}
Similarly, for the four-form we find
\begin{equation}
 \label{DeformedRR4FormT4}
 \begin{aligned}
  F_4 & =
  2 \m (1 - \l)^3 \sinh\at \cos\a_1 \Big( \frac{e^{0129}}{L_0} + \frac{e^{3459}}{L_1} \Big)
  \\[5pt] &
  + 2 \m (1 - \l) (1 + \l)^2 \cosh\at \sin\a_1 \Big( \frac{e^{2459}}{L_0} - \frac{e^{0139}}{L_1} \Big)
  \\[5pt] &
  + 2 \m (1 + \l) (1 - \l)^2 \Big(
  \frac{1}{L_0} \sinh\at \sin\a_1 \, e^{3678}
  - \frac{1}{L_1} \cosh\at \cos\a_1 \, e^{2678} \Big) \, ,
 \end{aligned}
\end{equation}
where for $e^6 , \, e^7$ and $e^8$ we consider \eqref{DeformedFrameT4}.

Notice that the background described above solves the type-IIA equations of motion, provided that $L_0 = L_1$, as can be inferred from \eqref{CondRadii} for large $L_2$.

\subsection{Solutions from other zoom-in limits}

It was shown in \cite{Sfetsos:2013wia} that when $\l$ approaches the identity, the $\l$-deformed model on a group reduces to the non-Abelian T-dual of the corresponding Principal Chiral Model (PCM). To ensure the consistency of the $\l \rightarrow 1$ limit one has to relate $\l$ to the WZW level $k$ in such a way that $\l \rightarrow 1$ as $k \rightarrow \infty$. At the same time, it is necessary to expand the group element near the identity by rescaling the group parameters appropriately with inverse powers of $k$.

At the level of the supergravity solution described in Section \ref{LambdaDeformedSolution}, the NATD limit is taken by setting
\begin{equation}
 \label{NATDlimit}
 \at = i \frac{\pi}{2} + \frac{r_0}{2k} \, , \qquad \a_1 = \frac{r_1}{2k} \, , \qquad \a_2 = \frac{r_2}{2 k} \, , \qquad \l = 1 - \frac{1}{k}
\end{equation}
and then sending $k$ to infinity. Notice that due to \eqref{FromSU2ToSL2R}, the $SL(2 , \mathbb{R})$ group element is not connected to the identity. Therefore, in order to define the NATD limit, one has to consider an expansion around a complex value of the $SL(2 , \mathbb{R})$ group parameters. As it has been observed in \cite{Itsios:2023kma}, this results in a background with real RR fluxes.

However, this is not the only consistent option one has as $\l$ approaches the identity. Alternatively, one can expand $\at$ near zero and the angles $\a_1$ and $\a_2$ near $\nicefrac{\pi}{2}$. This opens up seven additional possibilities for consideration. Nevertheless, we will not delve into all of these options and will instead concentrate solely on the non-Abelian T-dual limit. The corresponding field content is given below.

\noindent\textbf{The metric.} The line element of the NATD geometry reads
\begin{equation}
 \label{NATDmetric}
 \begin{aligned}
  ds^2 & = \frac{L^2_0}{2} \Big( dr^2_0 + \frac{r^2_0}{r^2_0 - 1} \big( d\bt^2 - \cosh^2\bt \, d\gt^2 \big) \Big)
  \\[5pt]
  & + \frac{L^2_1}{2} \Big( dr^2_1 + \frac{r^2_1}{r^2_1 + 1} \big( d\b^2_1 + \sin^2\b_1 \, d\g^2_1 \big) \Big)
  \\[5pt]
  & + \frac{L^2_2}{2} \Big( dr^2_2 + \frac{r^2_2}{r^2_2 + 1} \big( d\b^2_2 + \sin^2\b_2 \, d\g^2_2 \big) \Big) + d\om^2 \, .
 \end{aligned}
\end{equation}
Notice that in order to maintain the signature, $r_0$ is restricted as $|r_0| > 1$. Moreover, applying the limit \eqref{NATDlimit} in the frame \eqref{DeformedFrame} we find
\begin{equation}
 \begin{aligned}
  & \he^0 = \frac{L_0}{\sqrt{2}} \frac{r_0}{\sqrt{r^2_0 - 1}} \, \cosh \bt \, d\gt \, , \quad
      \he^1 = \frac{L_0}{\sqrt{2}} \frac{r_0}{\sqrt{r^2_0 - 1}} \, d\bt \, , \quad
      \he^2 = \frac{L_0}{\sqrt{2}} dr_0 \, ,
  \\[5pt]
  & \he^3 = \frac{L_1}{\sqrt{2}} dr_1 \, ,\quad
     \he^4 = \frac{L_1}{\sqrt{2}} \frac{r_1}{\sqrt{r^2_1 + 1}} \, d\b_1 \, , \quad
     \he^5 = \frac{L_1}{\sqrt{2}} \frac{r_1}{\sqrt{r^2_1 + 1}} \, \sin\b_1 \, d\g_1 \, ,
  \\[5pt]
  & \he^6 = \frac{L_2}{\sqrt{2}} dr_2 \, ,\quad
     \he^7 = \frac{L_2}{\sqrt{2}} \frac{r_2}{\sqrt{r^2_2 + 1}} \, d\b_2 \, , \quad
     \he^8 = \frac{L_2}{\sqrt{2}} \frac{r_2}{\sqrt{r^2_2 + 1}} \, \sin\b_2 \, d\g_2 \, , \quad
  \\[5pt]
  & \he^9 = d\om \, .
 \end{aligned}
\end{equation}

\noindent\textbf{The dilaton.} In the NATD limit \eqref{NATDlimit} the dilaton diverges. Nevertheless, the divergence can be absorbed shifting the dilaton by a constant that depends on $k$. As we will explain later, this shift will be taken into account when we consider the NATD limit in the RR forms. Coming back to the dilaton, this reduces to
\begin{equation}
 \label{NATDdilaton}
 \Phi = - \frac{1}{2} \ln \left( \big( r^2_0 - 1 \big) \big( r^2_1 + 1 \big) \big( r^2_2 + 1 \big) \right) \, .
\end{equation}

\noindent\textbf{The NS form.} The two-form \eqref{DeformedTwoForm} has a good behaviour under the limit \eqref{NATDlimit} where one finds
\begin{equation}
 \label{NATDtwoform}
 \begin{aligned}
  B_2 & = \frac{L^2_0}{2} \frac{r^3_0}{r^2_0 - 1} \cosh\bt \, d\bt \wedge d\gt
  - \frac{L^2_1}{2} \frac{r^3_1}{r^2_1 + 1} \sin\b_1 \, d\b_1 \wedge d\g_1
  \\[5pt] &
  - \frac{L^2_2}{2} \frac{r^3_2}{r^2_2 + 1} \sin\b_2 \, d\b_2 \wedge d\g_2 \, . \end{aligned}
\end{equation}

\noindent\textbf{The RR forms.} The shift in the dilaton that we mentioned earlier is not innocent when considering the NATD limit in the RR forms. Due to this, we need to rescale the RR fields by an appropriate $k$-dependent constant. This ensures that the supergravity equations of motion are satisfied after the limit \eqref{NATDlimit}. The resulting RR sector contains a two- and a four-form. The first is
\begin{equation}
 \label{NATDRRtwoform}
 F_2 = \sqrt{2} \Big( \frac{r_1 r_2}{L_0} \he^{36} - \frac{r_0 r_2}{L_1} \he^{26} + \frac{r_0 r_1}{L_2} \he^{23} - \frac{r_0}{L_0} \he^{29} + \frac{r_1}{L_1} \he^{39} + \frac{r_2}{L_2} \he^{69} \Big) \, .
\end{equation}
For the four-form we find
\begin{equation}
 \label{NATDRRfourform}
 \begin{aligned}
  F_4 = & - \sqrt{2} \, r_0 r_1 r_2 \Big( \frac{\he^{0136}}{L_0} - \frac{\he^{2456}}{L_1} + \frac{\he^{2378}}{L_2} \Big)
  + \sqrt{2} r_0 r_1 \Big( \frac{\he^{2459}}{L_0} - \frac{\he^{0139}}{L_1} \Big)
  \\[5pt] &
  + \sqrt{2} r_0 r_2 \Big( \frac{\he^{2789}}{L_0} - \frac{\he^{0169}}{L_2} \Big)
  - \sqrt{2} r_1 r_2 \Big( \frac{\he^{3789}}{L_1} + \frac{\he^{4569}}{L_2} \Big)
  \\[5pt] &
  - \sqrt{2} r_0 \Big( \frac{\he^{2678}}{L_1} - \frac{\he^{2345}}{L_2} \Big)
  + \sqrt{2} r_1 \Big( \frac{\he^{3678}}{L_0} - \frac{\he^{0123}}{L_2} \Big)
  + \sqrt{2} r_2 \Big( \frac{\he^{3456}}{L_0} + \frac{\he^{0126}}{L_1} \Big)
  \\[5pt] &
  + \sqrt{2} \Big( \frac{\he^{0129}}{L_0} + \frac{\he^{3459}}{L_1} + \frac{\he^{6789}}{L_2} \Big) \, .
 \end{aligned}
\end{equation}
Again, in order to guarantee that the supergravity equations of motion are satisfied we need to make use of \eqref{CondRadii}.

\subsection{Uplift to eleven dimensions}
\label{ElevenDimLambda}

We know that any type-IIA solution with zero Romans mass can be lifted to eleven dimensions. Supergravity in eleven dimensions contains only the metric, $ds^2_{11}$, and a four-form, $G_4$. The type-IIA fields are encoded in the eleven-dimensional ones in the following way
\begin{equation}
 \label{Uplift11d}
 \begin{aligned}
  & ds^2_{11} = e^{- \frac{2}{3} \Phi} ds^2_{IIA} + e^{\frac{4}{3} \Phi} \big( dx_{11} + C_1 \big)^2 \, ,
  \\[5pt]
  & G_4 = F_4 + H \wedge \big( dx_{11} + C_1 \big) \, .
 \end{aligned}
\end{equation}
In our case, $ds^2_{IIA}$ is the line element \eqref{DeformedMetric}, $\Phi$ is the dilaton \eqref{DeformedDilaton}, $H$ is the field strength of \eqref{DeformedTwoForm}, $F_4$ is given in \eqref{DeformedRR4Form} and $C_1$ is the RR potential of \eqref{DeformedRR2Form} such that $F_2 = dC_1$. For $C_1$ we find
\begin{equation}
 \begin{aligned}
  C_1 & = 2 \, \m \, k \, (1 + \l)^3 L_0 \sinh\at \Big( \frac{L_1}{L_2} \sin\a_1 \cos\a_2 \, d\a_1 - \frac{L_2}{L_1} \cos\a_1 \sin\a_2 \, d\a_2 \Big)
  \\[5pt] &
  - 2 \m \sqrt{k (1 - \l^2)^3} \sinh\at \cos\a_1 \cos\a_2 \, d\om \, .
 \end{aligned}
\end{equation}
To make sure that the eleven-dimensional equations of motion are satisfied it is necessary to take into account \eqref{CondRadii}.

\section{Supersymmetry analysis}
\label{SupersymmetryAnalysisSynopsis}

We will now provide a summary of the supersymmetry analysis for the type-IIA background described in equations \eqref{DeformedMetric}, \eqref{DeformedTwoForm}, \eqref{DeformedDilaton} \eqref{DeformedRR2Form} and \eqref{DeformedRR4Form}. Further details can be found in Appendix \ref{SupersymmetryAnalysis}. Due to the fact that the undeformed background exhibits more symmetry than the deformed one, it is instructive to differentiate between the two cases.

\noindent\textbf{The $\l = 0$ case.} As it is explained in Appendix \ref{SupersymmetryAnalysis} the dilatino equation is solved by imposing a single projection, namely the one outlined in \eqref{Projectionlambda0}. On the other hand, the gravitini variations are solved by the Killing spinor
\begin{equation}
 \begin{aligned}
  \e = & \exp \Big( - \frac{\at}{2} \G^{01} \s_3 \Big) \, \exp \Big( \frac{\bt}{2} \G^{02} \s_3 \Big) \, \Big( - \frac{\gt}{2} \G^{12} \s_3 \Big)
  \\[5pt]
  \times & \exp \Big( - \frac{\a_1}{2} \G^{45} \s_3 \Big) \, \exp \Big( \frac{\b_1}{2} \G^{34} \Big) \, \Big( \frac{\g_1}{2} \G^{45} \Big)
  \\[5pt]
  \times & \exp \Big( - \frac{\a_2}{2} \G^{78} \s_3 \Big) \, \exp \Big( \frac{\b_2}{2} \G^{67} \Big) \, \Big( \frac{\g_2}{2} \G^{78} \Big) \eta \, ,
 \end{aligned}
\end{equation}
where $\eta$ is a constant spinor subject to the projection
\begin{equation}
 \label{Projectionlambda0eta}
 \Big( \frac{L_0}{L_1} \, \G^{012345} + \frac{L_0}{L_2} \, \G^{012678} \Big) \eta = - \eta \, .
\end{equation}
Notice that \eqref{CondRadii} ensures that the operator inside the parenthesis squares to the identity. The condition \eqref{Projectionlambda0eta} implies that the undeformed solution preserves $16$ supercharges, in agreement with the previously known result \cite{Cowdall:1998bu}.

\noindent\textbf{The $\l \ne 0$ case.} Allowing for non-trivial values of the deformation parameter generates RR fluxes. The presence of the RR fields results in extra terms in the dilatino equation, which vanish by imposing an additional projection. On top of \eqref{Projectionlambda0}, eq. \eqref{Projectionlambdane0} needs to be considered. The gravitinii can be integrated by applying the type-IIA chirality condition in conjunction with the aforementioned projections. This yields an expression for the Killing spinor which now depends on $\l$
\begin{equation}
 \label{KillingSpinor}
 \begin{aligned}
  \e = & \exp \left( - \frac{1}{2} \tanh^{-1} \Big( \frac{1 - \l}{1 + \l} \tanh\at \Big) \G^{01} \s_3 \right) \, \exp \Big( \frac{\bt}{2} \G^{02} \s_3 \Big) \, \Big( - \frac{\gt}{2} \G^{12} \s_3 \Big)
  \\[5pt]
  \times & \exp \left( - \frac{1}{2} \tan^{-1} \Big( \frac{1 + \l}{1 - \l} \tan\a_1 \Big) \G^{45} \s_3 \right) \, \exp \Big( \frac{\b_1}{2} \G^{34} \Big) \, \Big( \frac{\g_1}{2} \G^{45} \Big)
  \\[5pt]
  \times & \exp \left( - \frac{1}{2} \tan^{-1} \Big( \frac{1 + \l}{1 - \l} \tan\a_2 \Big) \G^{78} \s_3 \right) \, \exp \Big( \frac{\b_2}{2} \G^{67} \Big) \, \Big( \frac{\g_2}{2} \G^{78} \Big) \eta \, .
 \end{aligned}
\end{equation}
Again, $\eta$ is a constant spinor which now satisfies
\begin{equation}
 \label{Projectionlambdane0eta}
 \G^{019} \s_1 \eta = - \eta
\end{equation}
together with the independent projection \eqref{Projectionlambda0eta}. The necessity to impose the second projection \eqref{Projectionlambdane0eta} when $\l \ne 0$ implies that the deformed solution preserves $8$ supercharges. In other words, the deformation breaks supersymmetry by half.

In the case of the NATD solution given in \eqref{NATDmetric}, \eqref{NATDdilaton}, \eqref{NATDtwoform}, \eqref{NATDRRtwoform} and \eqref{NATDRRfourform}, one can proceed by analysing the dilatino and gravitino variations directly or to simply take the limit \eqref{NATDlimit} in \eqref{KillingSpinor} and \eqref{Projectionlambdane0}. Either approach results in the Killing spinor

\begin{equation}
 \begin{aligned}
  \e = & \exp \Big( - \frac{1}{2} \coth^{-1} r_0 \, \G^{01} \s_3 \Big) \, \exp \Big( \frac{\bt}{2} \G^{02} \s_3 \Big) \, \Big( - \frac{\gt}{2} \G^{12} \s_3 \Big)
  \\[5pt]
  \times & \exp \Big( - \frac{1}{2} \tan^{-1} r_1 \, \G^{45} \s_3 \Big) \, \exp \Big( \frac{\b_1}{2} \G^{34} \Big) \, \Big( \frac{\g_1}{2} \G^{45} \Big)
  \\[5pt]
  \times & \exp \Big( - \frac{1}{2} \tan^{-1} r_2 \, \G^{78} \s_3 \Big) \, \exp \Big( \frac{\b_2}{2} \G^{67} \Big) \, \Big( \frac{\g_2}{2} \G^{78} \Big) \eta \, ,
 \end{aligned}
\end{equation}
with $\eta$ being a constant spinor satisfying \eqref{Projectionlambda0eta} and \eqref{Projectionlambdane0eta}. Therefore, the NATD solution also preserves $8$ supercharges.

\section{Penrose limits}
\label{PenroseLimits}

In this section, we explore an alternative zoom-in limit known as the Penrose limit \cite{Penrose1976}. This indicates that in the neighbourhood of a null geodesic the spacetime geometry resembles that of a plane-wave. Applying this to the $\l$-deformed background of eq. \eqref{DeformedMetric}, \eqref{DeformedTwoForm}, \eqref{DeformedDilaton} \eqref{DeformedRR2Form} and \eqref{DeformedRR4Form} results in another solution of the type-IIA supergravity. In the rest we derive the type-IIA solutions on the plane-wave geometries associated to the following two null geodesics
\begin{subequations}
 \begin{equation}
  \label{Motiongtg}
  \at = \bt = \a_2 = 0 \, , \qquad \a_1 = \b_1 = \frac{\pi}{2} \, ,
 \end{equation}
 \begin{equation}
  \label{Motiongtom}
  \at = \bt = \a_1 = \a_2 = 0 \, .
 \end{equation}
\end{subequations}
The first is related to the motion of a particle along the $U(1)$ directions $(\gt , \, \g_1)$ and the second along $(\gt , \, \om)$.

\subsection{The pp-wave around \eqref{Motiongtg}}

Let us start with the Penrose limit that corresponds to the null geodesic \eqref{Motiongtg}. This is obtained by setting
\begin{align}
  & \gt = \frac{u}{L_0} \, , \quad
     \g_1 = \frac{u}{L_1} + \frac{1 + \l}{1 - \l} \frac{v}{L_1 \, k} \, , \quad
      \at = \sqrt{\frac{1 - \l}{1 + \l}} \, \frac{z_1}{L_0 \sqrt{k}} \, , \quad
      \a_1 = \frac{\pi}{2} + \sqrt{\frac{1 - \l}{1 + \l}} \, \frac{z_2}{L_1 \sqrt{k}} \, ,
      \nonumber\\[5pt]
  & \a_2 = \sqrt{\frac{1 - \l}{1 + \l}} \, \frac{\r}{L_2 \sqrt{k}} \, , \quad 
     \bt = \sqrt{\frac{1 + \l}{1 - \l}} \, \frac{z_3}{L_0 \sqrt{k}} \, , \quad
     \b_1 = \frac{\pi}{2} + \sqrt{\frac{1 + \l}{1 - \l}} \, \frac{z_4}{L_1 \sqrt{k}}
\end{align}
and then taking $k$ to infinity. As a result we find the type-IIA solution with the following content
\begin{equation}
 \label{PlaneWaveMotiongtg}
 \begin{aligned}
  ds^2 & = 2 \, du dv + d\vec{z}^2_4 + d\vec{x}^2_4 + \cH \, du^2 \, , \quad \cH = \Bigg[ \Big( \frac{1 - \l}{1 + \l} \Big)^4 \Big( \frac{z^2_1}{L^2_0} + \frac{z^2_2}{L^2_1} \Big) + \frac{z^2_3}{L^2_0} + \frac{z^2_4}{L^2_1} \Bigg] \, ,
  \\[5pt]
  H_3 & = 2 \frac{1 + \l^2}{\big( 1 + \l \big)^2} du \wedge \Big( \frac{1}{L_0} dz_1 \wedge dz_3 - \frac{1}{L_1} dz_2 \wedge dz_4 \Big) \, , \qquad \quad
  \Phi = 0 \, ,
  \\[5pt]
  F_4 & = \frac{4 \l}{\big( 1 + \l \big)^2} \, du \wedge d\om \wedge \Big( \frac{1}{L_0} dz_1 \wedge dz_4 + \frac{1}{L_1} dz_2 \wedge dz_3 \Big) \, , 
 \end{aligned}
\end{equation}
where we represent the four-dimensional flat space transverse to the $u$, $v$ and $z$ directions as $d\vec{x}^2_4 = d\r^2 + \r^2 \big( d\b^2_2 + \sin^2\b_2 \, d\g^2_2 \big) + d\om^2$. Therefore, the line element above is manifestly in the Brinkmann form. Notice that we have set the dilaton to zero after a shift by a suitable constant. This shift must be accompanied by an appropriate rescaling of the RR four-form in order to guarantee that the type-IIA equations of motion are still satisfied. Taking $L_0 = L_1 = 1$ in \eqref{PlaneWaveMotiongtg}, it becomes evident that the plane-wave solution is T-dual to the one found in Section $3.2$ of \cite{Itsios:2023kma}.

\noindent\textbf{Uplift to eleven dimensions.} The plane-wave solution \eqref{PlaneWaveMotiongtg} can be lifted to eleven dimensions using \eqref{Uplift11d} where we find
\begin{equation}
 \label{PlaneWaveMotiongtg11D}
 \begin{aligned}
  ds^2 & = 2 \, du dv + d\vec{z}^2_4 + d\vec{x}^2_4 + dx^2_{11} + \cH \, du^2
  \\[5pt]
  G_4 & = \frac{4 \l}{\big( 1 + \l \big)^2} \, du \wedge d\om \wedge \Big( \frac{1}{L_0} dz_1 \wedge dz_4 + \frac{1}{L_1} dz_2 \wedge dz_3 \Big)
  \\[5pt]
  & + 2 \frac{1 + \l^2}{\big( 1 + \l \big)^2} du \wedge dx_{11} \wedge \Big( \frac{1}{L_0} dz_1 \wedge dz_3 - \frac{1}{L_1} dz_2 \wedge dz_4 \Big) \, .
 \end{aligned}
\end{equation}
Alternatively, this can be obtained by applying the Penrose limit for the null geodesic \eqref{Motiongtg} directly in the eleven-dimensional solution described in Section \ref{ElevenDimLambda}.

\noindent\textbf{Remarks on supersymmetry.} As previously mentioned, when $L_0 = L_1 = 1$ the plane-wave solution \eqref{PlaneWaveMotiongtg} is T-dual to the plane-wave solution found in \cite{Itsios:2023kma}. The latter has been shown to preserve $16$ supercharges when $\l \ne 0$, while for $\l = 0$, the existence of $8$ supernumerary supercharges ($24$ in total) has been observed. However, \eqref{PlaneWaveMotiongtg} results from a geometry with $L_0 \ne L_1$. This implies that even when $\l = 0$ \cite{Sadri:2003ib} the background \eqref{PlaneWaveMotiongtg} still preserves $16$ supercharges. This statement can be confirmed simply by looking at the dilatino equation, which reads
\footnote{
 The $\G$-matrices here are associated to the frame
 \begin{equation}
  \begin{aligned}
   & e^{+} = dv + \frac{1}{2} \cH \, du \, , \qquad e^{-} = du \, , \qquad
  e^0 = \frac{1}{\sqrt{2}} \big( e^{+} - e^{-} \big) \, , \quad e^9 = \frac{1}{\sqrt{2}} \big( e^{+} + e^{-} \big) \, ,
  \\[5pt]
  & e^1 = dz_1 \, , \quad e^2 = dz_2 \, , \quad e^3 = dz_3 \, , \quad e^4 = dz_4 \, , \quad
     e^5 = dx_1 \, , \quad e^6 = dx_2 \, , \quad e^7 = dx_3 \, , \quad e^8 = dx_4 \, .
  \end{aligned}
 \end{equation}
 }
\begin{equation}
 \d\l = - \frac{1}{2 L_0} \frac{1 + \l^2}{(1 + \l)^2} \G^{-} \G^{13} \s_3 \Big( \mathbb{1} + \frac{\l}{1 + \l^2} \G^{138} (i \s_2) \Big) \Big( \mathbb{1} - \frac{L_0}{L_1} \G^{1234} \Big) \e \, .
\end{equation}
Clearly, when $L_0 \ne L_1$, the vanishing of the dilatino variation implies that the spinor $\e$ is annihilated only by $\G^{-}$, and therefore, the pp-wave solution preserves $16$ supercharges.

For completeness, below we present the spinor $\e$ that solves the gravitino variations
\begin{equation}
 \begin{aligned}
  \e & = \exp \left( \frac{u}{2} \frac{1 + \l^2}{(1 + \l)^2} \Big( \frac{1}{L_0} \G^{13} - \frac{1}{L_1} \G^{24} \Big) \s_3 \right)
  \\[5pt]
  & \times \exp \left(\frac{u}{2} \frac{\l}{(1 + \l)^2} \G^{-} \G^{+} \G^8 \Big( \frac{1}{L_0} \G^{14} + \frac{1}{L_1} \G^{23} \Big) \s_1 \right) \eta \, ,
 \end{aligned}
\end{equation}
where $\eta$ is a constant spinor satisfying $\G^{-} \eta = 0$.

\subsection{The pp-wave around \eqref{Motiongtom}}

For the Penrose limit related to the null geodesic \eqref{Motiongtom} we set
\begin{align}
  & \gt = \frac{u}{L_0} \, , \qquad
     \om = \sqrt{k \frac{1 - \l}{1 + \l}} u + \sqrt{\frac{1 + \l}{1 - \l}} \frac{v}{\sqrt{k}} \, , \qquad
      \at = \sqrt{\frac{1 - \l}{1 + \l}} \, \frac{z_1}{L_0 \sqrt{k}} \, ,
      \nonumber\\[5pt]
  & \bt = \sqrt{\frac{1 + \l}{1 - \l}} \, \frac{z_2}{L_0 \sqrt{k}} \, , \qquad
     \a_1 = \sqrt{\frac{1 - \l}{1 + \l}} \, \frac{\r_1}{L_1 \sqrt{k}} \, , \qquad
     \a_2 = \sqrt{\frac{1 - \l}{1 + \l}} \, \frac{\r_2}{L_2 \sqrt{k}} \, .
\end{align}
Sending $k$ to infinity we obtain the plane-wave solution with content
\begin{equation}
 \label{PlaneWaveMotiongtom}
 \begin{aligned}
  ds^2 & = 2 \, du dv + dz^2_1 + dz^2_2 + d\vec{y}^2_3 + d\vec{w}^2_3 + \widetilde{\cH} \, du^2 \, , \quad \widetilde{\cH} = - \Bigg[ \Big( \frac{1 - \l}{1 + \l} \Big)^4 \, \frac{z^2_1}{L^2_0} + \frac{z^2_2}{L^2_0} \Bigg] \, ,
  \\[5pt]
  \Phi & = 0 \, , \qquad
  H_3 = \frac{2}{L_0} \frac{1 + \l^2}{\big( 1 + \l \big)^2} \, du \wedge dz_1 \wedge dz_2 \, , \qquad
  F_2 = \frac{4 \, \l}{L_0 \big( 1 + \l \big)^2} \, du \wedge dz_1 \, .
 \end{aligned}
\end{equation}
Here, $d\vec{y}^2_3$ and $d\vec{w}^2_3$ stand for the line elements of two three-dimensional flat spaces. In particular, $d\vec{y}^2_3 = d\r^2_1 + \r^2_1 \, \big( d\b^2_1 + \sin^2\b_1 \, d\g^2_1 \big)$ and $d\vec{w}^2_3 = d\r^2_2 + \r^2_2 \, \big( d\b^2_2 + \sin^2\b_2 \, d\g^2_2 \big)$. Clearly, the line element in \eqref{PlaneWaveMotiongtom} is in Brinkmann form. Notice that again we have set the dilaton to zero by a suitable shift and rescaled the RR two-form accordingly, as we did for \eqref{PlaneWaveMotiongtg}.

\noindent\textbf{Uplift to eleven dimensions.} Unlike \eqref{PlaneWaveMotiongtg}, in \eqref{PlaneWaveMotiongtom} the RR two-form survived the Penrose limit. Its potential signals a fibration term along the extra coordinate after lifting the metric in \eqref{PlaneWaveMotiongtom} to eleven dimensions. In particular, for the eleven-dimensional solution we find
\begin{equation}
 \label{PlaneWaveMotiongtom11D}
 \begin{aligned}
  ds^2 = & 2 \, du dv + dz^2_1 + dz^2_2 + d\vec{y}^2_3 + d\vec{w}^2_3
  + \left( dx_{11} + \widetilde{C}_{1} du \right)^2
  + \widetilde{\cH} \, du^2 \, ,
  \\[5pt]
  \widetilde{C}_{1}= & - \frac{4 \, \l \, z_1}{L_0 \big( 1 + \l \big)^2} \, du \, , \qquad
  G_4 = \frac{2}{L_0} \, \frac{1 + \l^2}{\big( 1 + \l \big)^2} \, du \wedge dx_{11} \wedge dz_1 \wedge dz_2 \, .
 \end{aligned}
\end{equation}
The above content can also be found by applying the Penrose limit around the null geodesic \eqref{Motiongtom} directly in the eleven-dimensional background of Section \ref{ElevenDimLambda}. 

Notice that due to the fibration term along $x_{11}$, the line element in \eqref{PlaneWaveMotiongtom11D} is not in Brinkmann form. In order to bring it to the Brinkmann form we apply the following coordinate transformations
\begin{equation}
 \label{ToBrinkmann}
 v \mapsto v + \frac{2 \, \l \, z_1 \, x_{11}}{L_0 \big( 1 + \l \big)^2} \, , \qquad
 x_{11} + i \, z_1 \mapsto e^{- i \tilde{u} }\big( x_{11} + i \, z_1 \big) \, , \qquad
 \tilde{u} = \frac{2 \, \l \, u}{L_0 \big( 1 + \l \big)^2} \, .
\end{equation}
As a result, the fibration disappears but the mass term (coefficient of $du^2$) acquires dependence on $u$, namely
\begin{equation}
 \begin{aligned}
  ds^2 & = 2 \, du dv + dz^2_1 + dz^2_2 + d\vec{y}^2_3 + d\vec{w}^2_3
  - \Bigg( \frac{4 \, \l^2}{L^2_0 \big( 1 + \l \big)^4} \big( x^2_{11} + z^2_1 \big) + \frac{z^2_2}{L^2_0}
  \\[5pt]
  & + \frac{1 - 6 \l + \l^2}{L^2_0 \big( 1 + \l \big)^2} \big( x_{11} \sin \tilde{u} - z_1 \cos \tilde{u} \big)^2 \Bigg) du^2 \, .
 \end{aligned}
\end{equation}
The four-form $G_4$ remains invariant under the transformation \eqref{ToBrinkmann}.

\noindent\textbf{Remarks on supersymmetry.} Looking at the dilatino variation for the plane-wave background \eqref{PlaneWaveMotiongtom}, one might anticipate that there is room for supernumerary supercharges. Indeed, the dilatino variation in this case takes the form
\footnote{
 The $\G$-matrices here are associated to the frame
 \begin{align}
   & e^{+} = dv + \frac{1}{2} \widetilde{\cH} \, du \, , \qquad e^{-} = du \, , \qquad
  e^0 = \frac{1}{\sqrt{2}} \big( e^{+} - e^{-} \big) \, , \quad e^9 = \frac{1}{\sqrt{2}} \big( e^{+} + e^{-} \big) \, ,
  \\[5pt]
  & e^1 = dz_1 \, , \quad e^2 = dz_2 \, , \quad e^3 = dy_1 \, , \quad e^4 = dy_2 \, , \quad
     e^5 = dy_3 \, , \quad e^6 = dw_1 \, , \quad e^7 = dw_2 \, , \quad e^8 = dw_3 \, .
     \nonumber
 \end{align}
 }
\begin{equation}
 \d\l = - \frac{1}{2 L_0} \frac{1 + \l^2}{(1 + \l)^2} \G^{-} \G^{12} \s_3 \Big( \mathbb{1} - \frac{3 \l}{1 + \l^2} \G^2 \s_1 \Big) \e \, .
\end{equation}
It is now clear that the term in the parenthesis defines a projector when $\l = \nicefrac{(3 - \sqrt{5})}{2}$. In other words, for this value of $\l$, one should expect the existence of $8$ supernumerary supercharges. Nevertheless, this turns out not to be true after elaborating on the gravitino variations. In fact, the gravitini imply that the spinor $\e$ is subject to the conditions
\begin{equation}
 \begin{aligned}
  & \l \G^{-} \G^1 \Big( 4 - 3 \l + 4 \l^2 + 2 \big( 1 + \l^2 \big) \G^2 \s_1 \Big) \e = 0 \, ,
  \\[5pt]
  & \l \G^{-} \G^2 \Big( 4 + 3 \l + 4 \l^2 + 2 \big( 1 + \l^2 \big) \G^2 \s_1 \Big) \e = 0 \, ,
  \\[5pt]
  & \l^2 \, \G^{-} \G^i \e = 0 \, , \qquad i = 3 , \ldots , 8 \, .
 \end{aligned}
\end{equation}
The above relations are simultaneously satisfied only when $\G^{-} \e = 0$. This means that the plane-wave solution \eqref{PlaneWaveMotiongtom} preserves the minimum amount of supersymmetry. 

Let us mention that after integrating the gravitino equations we find that the spinor $\e$ takes the form
\begin{equation}
 \e = \exp\Bigg( \frac{u}{2 L_0} \frac{1 + \l^2}{(1 + \l)^2} \G^{12} \s_3 + \frac{u}{2 L_0} \frac{\l}{(1 + \l)^2} \G^{-} \G^{+} \G^1 (i \s_2) \Bigg) \eta \, ,
\end{equation}
where $\eta$ is a constant spinor annihilated by $\G^{-}$.

\section{Conclusions}
\label{ConclusionsAndFutureIdeas}

We promoted the integrable $\l$-model on $SL(2 , \mathbb{R}) \times SU(2) \times SU(2)$ to a solution of type-IIA supergravity by constructing the appropriate RR fluxes. In the undeformed limit, we recover the geometry of $AdS_3 \times S^3 \times S^3 \times S^1$ supported only by a NS three-form. This background originates from the near-horizon limit of the NS1-NS5-NS5 brane intersection and is 1/2 supersymmetric. In the presence of the deformation, the group of isometries of $AdS_3 \times S^3 \times S^3$ reduces to that of $AdS_2 \times S^2 \times S^2$. As a matter of fact, the deformation breaks supersymmetry by half. As the deformation parameter approaches the identity, one is forced to employ a zoom-in limit, which involves rescaling the coordinates by factors of $k$ and sending $k$ to infinity. We demonstrated that there exist several such zoom-in limits and presented the one associated with the NATD of the type-IIB pure RR background on $AdS_3 \times S^3 \times S^3 \times S^1$. Finally, we considered Penrose limits along two null geodesics in the deformed spacetime. The resulting type-IIA plane-wave backgrounds capture a dependence on the deformation parameter. By analysing the supersymmetry variations for the dilatino and gravitino, we concluded that no supernumerary supercharges exist for arbitrary values of the deformation parameter.

Having a supergravity background with the aforementioned properties at hand opens the door for further investigations. Some of the questions that we would like to address in the near future are the following. As a first thought, it would be interesting to explore the existence of supersymmetric probe brane embeddings \cite{Cederwall:1996ri,Bergshoeff:1996tu,Arean:2004mm} by analysing the $\kappa$-symmetry condition. Furthermore, the appearance of an $AdS_2$ subspace in the deformed geometry suggests the study of the holographic dual to our solution as a natural extension. To this, one can also include the study of thermal effects. This requires to construct the corresponding black hole solution, which is equivalent to turning on temperature in the dual holographic system.

\section*{Acknowledgements}

I would like to thank K. Sfetsos for inspiring discussions and the Physics Department of the Aristotle University of Thessaloniki (Greece) for hospitality. This research is supported by the Einstein Stiftung Berlin via the Einstein International Postdoctoral Fellowship program ``Generalised dualities and their holographic applications to condensed matter physics'' (project number IPF- 2020-604). I would also like to acknowledge support by the Deutsche Forschungsgemeinschaft (DFG, German Research Foundation) via the Emmy Noether program ``Exploring the landscape of string theory flux vacua using exceptional field theory'' (project number 426510644).

\appendix

\section{The $\l$-deformed background fields}
\label{LambdaBackgroundFields}

In this Appendix we summarise the background fields for the $\l$-deformed $\s$-models on $SL(2, \mathbb{R})$ and $SU(2)$. These consist of a three-dimensional metric, a two-form and a scalar.

\noindent\textbf{The $SU(2)$ fields.} Starting with the compact case, the $\s$-model background fields are
\begin{equation}
 \begin{aligned}
  \label{SU2LambdaBackgroundFields}
  & ds^2 = k \Big( \frac{1 + \l}{1 - \l} \, d\a^2 + \frac{1 - \l^2}{\D(\a)} \sin^2\a \big( d\b^2 + \sin^2\b \, d\g^2 \big) \Big) \, ,
  \\[5pt]
  & B_2 = k \Big( - \a + \frac{(1 - \l)^2}{\D(\a)} \cos\a \, \sin\a \Big) \sin\b \, d\b \wedge d\g \, ,
  \\[5pt]
  & \Phi = - \frac{1}{2} \ln \D(\a) \, ,
 \end{aligned}
\end{equation}
where the function $\D$ is defined as
\begin{equation}
 \label{Delta}
 \D(x) := (1 - \l)^2 \, \cos^2 x + (1 + \l)^2 \, \sin^2 x \, .
\end{equation}
Notice that in the neighbourhood of the points $\a = 0 , \pi$ and $\a = \nicefrac{\pi}{2}$ the topology looks different. Specifically, near $\a = 0 , \pi$ it looks like $\mathbb{R}^3$ while near $\a = \nicefrac{\pi}{2}$ it behaves like $\mathbb{R} \times S^2$.

\noindent\textbf{The $SL(2, \mathbb{R})$ fields.} The corresponding fields for the non-compact case can be obtained from the above via the analytic continuation
\begin{equation}
 \label{FromSU2ToSL2R}
 \a \mapsto \frac{\pi}{2} + i \at \, , \qquad \b \mapsto i \bt - \frac{\pi}{2} \, , \qquad \g \mapsto \gt \, , \qquad k \mapsto - k \, .
\end{equation}
Doing so, one finds
\begin{equation}
 \begin{aligned}
  \label{SL2RLambdaBackgroundFields}
  & ds^2 = k \Big( \frac{1 + \l}{1 - \l} \, d\at^2 + \frac{1 - \l^2}{\tilde{\D}(\at)} \cosh^2\at \big( d\bt^2 - \cosh^2\bt \, d\gt^2 \big) \Big) \, ,
  \\[5pt]
  & B_2 = k \Big( \at + \frac{(1 - \l)^2}{\tilde{\D}(\at)} \cosh\at \, \sinh\at \Big) \cosh\bt \, d\bt \wedge d\gt \, ,
  \\[5pt]
  & \Phi = - \frac{1}{2} \ln \tilde{\D}(\at) \, ,
 \end{aligned}
\end{equation}
where the function $\tilde{\D}$ is given by
\begin{equation}
 \label{DeltaTilde}
 \tilde{\D}(x) := (1 + \l)^2 \cosh^2 x - (1 - \l)^2 \sinh^2 x \, .
\end{equation}
Notice that we have neglected an imaginary term in the two-form \eqref{SL2RLambdaBackgroundFields} that comes from the shift of $\a$ by $\nicefrac{\pi}{2}$. This term is a closed two-form and thus it is irrelevant for a supergravity solution. The line element \eqref{SL2RLambdaBackgroundFields} has mostly plus signature and the time-like direction is $\gt$. Moreover, when $\at$ approaches $\pm \infty$ the space looks like $\mathbb{R} \times AdS_2$.

\section{Type-II solutions on $AdS_3 \times S^3 \times S^3 \times S^1$}
\label{UndeformedSolutions}

Here we summarise the pure NS and RR (type-IIA/IIB) supergravity solutions with $AdS_3 \times S^3 \times S^3 \times S^1$ geometry. All of them have a trivial dilaton and a metric given in terms of the line element
%
 \begin{align}
  ds^2 & = L^2_0 \Big( d\at^2 + \cosh^2\at \big( d\bt^2 - \cosh^2\bt d\gt^2 \big) \Big) 
  + L^2_1 \Big( d\a^2_1 + \sin^2\a_1 \big( d\b^2_1 + \sin^2\b_1 d\g^2_1 \big) \Big)
  \nonumber\\[5pt]
  & + L^2_2 \Big( d\a^2_2 + \sin^2\a_2 \big( d\b^2_2 + \sin^2\b_1 d\g^2_2 \big) \Big) + d\om^2 \, .
  \label{UndeformedMetric}
 \end{align}
%
The $AdS_3$ space with radius $L_0$ is parametrised by $(\at , \bt , \gt)$, while the two round three-spheres with radii $L_1$ and $L_2$ are parametrised by $(\a_1 , \b_1 , \g_1)$ and $(\a_2 , \b_2 , \g_2)$ respectively. The $U(1)$ isometry that corresponds to shifts in the $\om$ direction represents the circle $S^1$. In order to present the various form fields we introduce the frame
\begin{equation}
 \label{UndeformedFrame}
 \begin{aligned}
  & \fe^0 = L_0 \cosh\at \cosh\bt \, d\gt \, , \qquad
     \fe^1 = L_0 \cosh\at \, d\bt \, , \qquad
     \fe^2 = L_0 \, d\at \, ,
  \\[5pt]
  & \fe^3 = L_1 \, d\a_1 \, , \qquad
     \fe^4 = L_1 \sin\a_1 \, d\b_1 \, , \qquad
     \fe^5 = L_1 \sin\a_1 \sin\b_1 \, d\g_1 \, ,
  \\[5pt]
  & \fe^6 = L_2 \, d\a_2 \, , \qquad
     \fe^7 = L_2 \sin\a_2 \, d\b_2 \, , \qquad
     \fe^8 = L_2 \sin\a_2 \sin\b_2 \, d\g_2 \, , \qquad
     \fe^9 = d\om \, .
 \end{aligned}
\end{equation}
The non-trivial form fields for each type of solutions are given below.

\noindent\textbf{The pure NS case.} This solution has only a NS three-form which is written in terms of the volume forms on $AdS_3$ and the two three-spheres as
\begin{equation}
 \label{UndeformedNSform}
 H_3 = - 2 \Big( \frac{1}{L_0} \, \fe^0 \wedge \fe^1 \wedge \fe^2 + \frac{1}{L_1} \, \fe^3 \wedge \fe^4 \wedge \fe^5 + \frac{1}{L_2} \, \fe^6 \wedge \fe^7 \wedge \fe^8 \Big) \, .
\end{equation}

\noindent\textbf{The pure RR type-IIB case.} The only non-trivial RR field in this background is the three-form which can also be written in terms of the volume forms on the two three-spheres as
\begin{equation}
 F_3 = 2 \Big(\frac{1}{L_0} \, \fe^0 \wedge \fe^1 \wedge \fe^2 + \frac{1}{L_1} \, \fe^3 \wedge \fe^4 \wedge \fe^5 + \frac{1}{L_2} \, \fe^6 \wedge \fe^7 \wedge \fe^8 \Big) \, .
\end{equation}

\noindent\textbf{The pure RR type-IIA case.} This is related to the previous type-IIB solution via T-duality in the $\om$ direction. The T-duality leaves the metric invariant and generates the four-form
\begin{equation}
 F_4 = 2 \Big( \frac{1}{L_0} \, \fe^0 \wedge \fe^1 \wedge \fe^2 + \frac{1}{L_1} \, \fe^3 \wedge \fe^4 \wedge \fe^5 + \frac{1}{L_2} \, \fe^6 \wedge \fe^7 \wedge \fe^8 \Big) \wedge \fe^9 \, .
\end{equation}

Notice that the above fields solve the equations of motion for type-II supergravity theories provided that the radii satisfy the following relation:
\begin{equation}
 \label{CondRadii}
 \frac{1}{L^2_0} = \frac{1}{L^2_1} + \frac{1}{L^2_2} \, .
\end{equation}

It is worth it to point out that one can recover the pure NS and RR (type-IIA/IIB) solutions on $AdS_3 \times S^3 \times T^4$ by employing the limit
\begin{equation}
\label{FromS3toT4}
 \a_2 = \frac{\r}{L_2} \, , \qquad L_2 \rightarrow \infty
\end{equation}
in the supergravity fields above. In that case, eq. \eqref{CondRadii} implies that $L_0 = L_1 = L$. In the same way, one could set $\a_1 = \nicefrac{\r}{L_1}$ and take $L_1 \rightarrow \infty$. The two limits are equivalent due to the symmetry of the supergravity solution on $AdS_3 \times S^3 \times S^3 \times S^1$, which allows for the exchange of the two three-spheres.

\section{The type-IIB deformed solution}
\label{TypeIIBsolution}

Starting with the type-IIA pure RR background on $AdS_3 \times S^3 \times S^3 \times S^1$ one can construct a $\l$-deformed type-IIB solution following the procedure described in Section \ref{LambdaDeformedSolution}. The NS sector of this theory is given by \eqref{DeformedMetric}, \eqref{DeformedTwoForm} and \eqref{DeformedDilaton}.

 Moving to the type-IIB RR forms we find that the deformation switches on all of them. More explicitly, the $F_1$ reads
\begin{equation}
 \label{DeformedRR1Form}
 \begin{aligned}
  F_1 = 2 \m (1 + \l) (1 - \l)^2 \Big( & \frac{1}{L_0} \cosh\at \cos\a_1 \cos\a_2 \, e^2 - \frac{1}{L_1} \sinh\at \sin\a_1 \cos\a_2 \, e^3 
  \\[5pt]
  & - \frac{1}{L_2} \sinh\at \cos\a_1 \sin\a_2 \, e^6 \Big) \, .
 \end{aligned}
\end{equation}
For the $F_3$ we find
\begin{equation}
 \label{DeformedRR3Form}
 \begin{aligned}
  F_3 = & - 2 \m (1 - \l) (1 + \l)^2 \left( \cosh\at \sin\a_1 \cos\a_2 \Big( \frac{e^{245}}{L_0} - \frac{e^{013}}{L_1} + \frac{e^{239}}{L_2} \Big) \right.
  \\[5pt] &
  + \cosh\at \cos\a_1 \sin\a_2 \Big( \frac{e^{278}}{L_0} - \frac{e^{269}}{L_1} - \frac{e^{016}}{L_2} \Big)
  \\[5pt] &
  \left. + \sinh\at \sin\a_1 \sin\a_2 \Big( \frac{e^{369}}{L_0} - \frac{e^{378}}{L_1} - \frac{e^{456}}{L_2} \Big) \right)
  \\[5pt] &
  - 2 \m (1 - \l)^3 \sinh\at \cos\a_1 \cos\a_2 \Big( \frac{e^{012}}{L_0} + \frac{e^{345}}{L_1} + \frac{e^{678}}{L_2} \Big) \, .
 \end{aligned}
\end{equation}
The self-dual form is $F_5 = (1 + \star) f_5$ where

 \begin{align}
   f_5 & = 2 \m (1 + \l)^3 \cosh\at \sin\a_1 \sin\a_2 \Big( \frac{e^{01369}}{L_0} - \frac{e^{01378}}{L_1} - \frac{e^{01456}}{L_2} \Big)
   \nonumber\\[5pt] &
   + 2 \m (1 + \l) (1 - \l)^2 \left( \sinh\at \sin\a_1 \cos\a_2 \Big( \frac{e^{01245}}{L_0} + \frac{e^{01239}}{L_2} \Big) \right.
   \label{DeformedRR5Form}\\[5pt] &
   + \sinh\at \cos\a_1 \sin\a_2 \Big( \frac{e^{01278}}{L_0} - \frac{e^{01269}}{L_1} \Big)
   \left. + \cosh\at \cos\a_1 \cos\a_2 \Big( \frac{e^{01345}}{L_1} + \frac{e^{01678}}{L_2} \Big) \right) \, .
   \nonumber
 \end{align}

Again, in order to guarantee that the above content solves the type-IIB equations of motion one needs to impose the condition \eqref{CondRadii}.

\noindent\textbf{The $T^4$ limit.} In the limit \eqref{FromS3toT4}, the background described above reduces to a solution of the type-IIB supergravity on the $\l$-deformed $AdS_3 \times S^3 \times T^4$ geometry. This has a NS sector which is summarised in \eqref{DeformedT4Metric}, \eqref{DeformedT4Dilaton} and \eqref{DeformedTwoFormT4}. The RR sector is obtained by applying \eqref{FromS3toT4} in \eqref{DeformedRR1Form}, \eqref{DeformedRR3Form} and \eqref{DeformedRR5Form}. It is easy to confirm that
%
 \begin{align}
  F_1 = & 2 \m (1 + \l) (1 - \l)^2 \Big( \frac{1}{L_0} \cosh\at \cos\a_1 \, e^2 - \frac{1}{L_1} \sinh\at \sin\a_1 \, e^3  \Big) \, ,
  \nonumber\\[5pt]
  F_3 = & - 2 \m (1 - \l) (1 + \l)^2 \cosh\at \sin\a_1 \Big( \frac{e^{245}}{L_0} - \frac{e^{013}}{L_1} \Big)
  \nonumber\\[5pt] &
  - 2 \m (1 - \l)^3 \sinh\at \cos\a_1 \Big( \frac{e^{012}}{L_0} + \frac{e^{345}}{L_1} \Big) \, ,
  \\[5pt]
  F_5 = & 2 \m (1 + \l) (1 - \l)^2 (1 + \star) \left( \frac{1}{L_0} \sinh\at \sin\a_1 e^{01245}
  \nonumber+ \frac{1}{L_1} \cosh\at \cos\a_1 e^{01345} \right) \, .
 \end{align}
%
The above content solves the type-IIB equations of motion, provided that $L_0 = L_1$, as understood from \eqref{CondRadii} for large $L_2$.

Notice that the above solution is the same as the one found in \cite{Itsios:2023kma} up to a shift in the dilaton and an appropriate rescaling of the RR forms. Moreover, it can be obtained from the background presented in Section \ref{T4Limit} by T-duality in the $e^9$ direction.

\section{Details on supersymmetry}
\label{DetailsOnSupersymmetry}

In this Appendix, we elaborate on the supersymmetry analysis for the type-IIA $\l$-deformed solution constructed in Section \ref{LambdaDeformedSolution}. We start by outlining the conventions for supersymmetry in type-IIA supergravity.

\subsection{Conventions}

For the variations under supersymmetry transformations of the dilatino and gravitino, we adopt the conventions from \cite{Hassan:1999bv}. In particular
\begin{equation}
 \begin{aligned}
  & \d\l = \frac{1}{2} \slashed{\partial} \Phi \e - \frac{1}{24} \slashed{H} \s_3 \e + \frac{e^{\Phi}}{8} \Big( 5 F_0 \s_1 + \frac{3}{2} \slashed{F}_2 \big( i \s_2 \big) + \frac{1}{24} \slashed{F}_4 \s_1 \Big) \e \, ,
  \\[5pt]
  & \d\psi_\m = D_\m \e - \frac{1}{8} H_{\m\n\r} \G^{\n\r} \s_3 \e + \frac{e^{\Phi}}{8} \Big( F_0 \s_1 + \frac{1}{2} \slashed{F_2} \big( i \s_2 \big) + \frac{1}{24} \slashed{F}_4 \s_1 \Big) \G_\m \e \, ,
 \end{aligned}
\end{equation}
where for completeness we also included the Romans mass $F_0$. The slash notation stands for contraction of the spacetime indices with antisymmetric products of $\G$-matrices. More precisely, we take $\slashed{\partial} = \G^\m \partial_\m$ and $\slashed{A}_p = A_{\m_1 \ldots \m_p} \G^{\m_1 \ldots \m_p}$. Moreover, $\s_i \, (i = 1,2,3)$ are the Pauli matrices and $\e$ represents a doublet
\begin{equation}
 \e =
 \begin{pmatrix}
  \e_+
  \\[5pt]
  \e_-
 \end{pmatrix}
\end{equation}
of two Majorana-Weyl spinors. In type-IIA supergravity, $\e$ satisfies the chirality condition $\G^{0 \ldots 9} \e = - \s_3 \e$, where the indices of the $\G$-matrices are associated with the Lorentzian tangent frame. It is understood that the Pauli matrices have a $GL(2)$ action on the doublet and for convenience we suppress the $GL(2)$ indices. Finally, $D_\m$ is the covariant derivative operator
\begin{equation}
 D_\m = \partial_\m + \frac{1}{4} \om_{\m a b} \G^{ab} \, , \qquad \om_{\m a b} = - \om_{\m b a} \, ,
\end{equation}
with $\om_{\m a b}$ being the spin-connection and we used Latin letters for the tangent frame indices.

\subsection{Killing spinor equations}
\label{SupersymmetryAnalysis}

We proceed with examining the supersymmetry of the background given in \eqref{DeformedMetric}, \eqref{DeformedTwoForm}, \eqref{DeformedDilaton} \eqref{DeformedRR2Form} and \eqref{DeformedRR4Form}. We begin with the dilatino variation which becomes notably simpler when we apply the projection
\begin{equation}
 \label{Projectionlambda0}
 \mathbb{P}_1 \e = - \e \, , \qquad \mathbb{P}_1 = \frac{L_0}{L_1} \, \G^{012345} + \frac{L_0}{L_2} \, \G^{012678} \, .
\end{equation}
Indeed, $\mathbb{P}_1$ squares to the identity in view of the restriction \eqref{CondRadii}. Moreover, it can be easily verified that the above projection solves the dilatino equation for $\l = 0$. However, when $\l \ne 0$, a lengthy calculation results in the following expression
\begin{align}
 \d\l = - \m (1 - \l) \Bigg( & \frac{\sinh\at}{L_0 \sqrt{\Dt(\at)}} \Big( \s_1 P_{\at} + P_{\a_1} (i \s_2) P_{\a_2} \Big)
 + \frac{\cos\a_1}{L_1 \sqrt{\D(\a_1)}} \Big( \s_1 P_{\a_1} - P_{\at} (i \s_2) P_{\a_2} \Big)
 \nonumber \\[5pt]
 & + \frac{\cos\a_2}{L_2 \sqrt{\D(\a_2)}} \Big( \s_1 P_{\a_2} + P_{\at} (i \s_2) P_{\a_1} \Big) \Bigg) \e \, .
\end{align}
In order to derive the above we made use of the projection condition \eqref{Projectionlambda0} and we defined the matrices
\begin{equation}
 \begin{aligned}
  & P_{\at} = \frac{(1 + \l) \cosh\at \, \G^2 \s_1 - (1 - \l) \sinh\at \, \G^{012} (i \s_2)}{\sqrt{\Dt(\at)}} \, ,
  \\[5pt]
  & P_{\a_1} = \frac{(1 + \l) \sin\a_1 \, \G^3 \s_1 + (1 - \l) \cos\a_1 \, \G^{345} (i \s_2)}{\sqrt{\D(\a_1)}} \, ,
  \\[7pt]
  & P_{\a_2} = \frac{(1 + \l) \sin\a_2 \, \G^6 \s_1 + (1 - \l) \cos\a_2 \, \G^{678} (i \s_2)}{\sqrt{\D(\a_2)}} \, .
 \end{aligned}
\end{equation}
It is now obvious that the dilatino equation is solved by imposing a second projection condition, namely
\begin{equation}
 \label{Projectionlambdane0}
 \mathbb{P}_2 \e = - \e \, , \qquad \mathbb{P}_2 = P_{\at} \s_1 P_{\a_1} (i \s_2) P_{\a_2} \, .
\end{equation}
Indeed, using the fact that $P^2_{\at} = P^2_{\a_1} = P^2_{\a_2} = \mathbb{1}$ and the commutation relations of $( P_{\at} , \, P_{\a_1} , \, P_{\a_2})$ with the Pauli matrices, one can show that $\mathbb{P}_2$ also squares to the identity and \eqref{Projectionlambdane0} implies
\begin{equation}
 \Big( \s_1 P_{\at} + P_{\a_1} (i \s_2) P_{\a_2} \Big) \e = \Big( \s_1 P_{\a_1} - P_{\at} (i \s_2) P_{\a_2} \Big) \e = \Big( \s_1 P_{\a_2} + P_{\at} (i \s_2) P_{\a_1} \Big) \e = 0 \, .
\end{equation}

Turning to the gravitino variation, we provide simplified expressions for its components upon imposing the type-IIA chirality condition and the projections \eqref{Projectionlambda0} and \eqref{Projectionlambdane0}. In particular, the non-trivial ones read

\noindent\textbf{The $\at$ component}
\begin{equation}
 \label{atcomponent}
 0 = \partial_{\at} \e + \frac{1 - \l^2}{2 \, \Dt(\at)} \G^{01} \s_3 \e \, .
\end{equation}

\noindent\textbf{The $\bt$ component}
\begin{equation}
 \label{btcomponent}
 0 = \partial_{\bt} \e - \frac{1}{2} \G^0 (i \s_2) P_{\at} \e \, .
\end{equation}

\noindent\textbf{The $\gt$ component}
\begin{equation}
 \label{gtcomponent}
 0 = \partial_{\gt} \e - \frac{1}{2} \sinh\bt \, \G^{01} \e + \frac{1}{2} \cosh\bt \, \G^1 (i \s_2) P_{\at} \e \, .
\end{equation}

\noindent\textbf{The $\a_1$ component}
\begin{equation}
 \label{a1component}
 0 = \partial_{\a_1} \e + \frac{1 - \l^2}{2 \, \D(\a_1)} \G^{45} \s_3 \e \, .
\end{equation}

\noindent\textbf{The $\b_1$ component}
\begin{equation}
 \label{b1component}
 0 = \partial_{\b_1} \e - \frac{1}{2} P_{\a_1} \G^5 (i \s_2) \e \, .
\end{equation}

\noindent\textbf{The $\g_1$ component}
\begin{equation}
 \label{g1component}
 0 = \partial_{\g_1} \e - \frac{1}{2} \cos\b_1 \, \G^{45} \e - \frac{1}{2} \sin\b_1 \, \G^4 (i \s_2) P_{\a_1} \e \, .
\end{equation}

\noindent\textbf{The $\a_2$ component}
\begin{equation}
 \label{a2component}
 0 = \partial_{\a_2} \e + \frac{1 - \l^2}{2 \, \D(\a_2)} \G^{78} \s_3 \e \, .
\end{equation}

\noindent\textbf{The $\b_2$ component}
\begin{equation}
 \label{b2component}
 0 = \partial_{\b_2} \e - \frac{1}{2} P_{\a_2} \G^8 (i \s_2) \e \, .
\end{equation}

\noindent\textbf{The $\g_2$ component}
\begin{equation}
 \label{g2component}
 0 = \partial_{\g_2} \e - \frac{1}{2} \cos\b_2 \, \G^{78} \e - \frac{1}{2} \sin\b_2 \, \G^7 (i \s_2) P_{\a_2} \e \, .
\end{equation}

It is evident that the subsets of equations for $(\at , \, \bt , \, \gt)$, $(\a_1 , \, \b_1 , \, \g_1)$ and $(\a_2 , \, \b_2 , \, \g_2)$ can be solved independently as they do not mix. Starting with the $\at$ component \eqref{atcomponent}, a straightforward integration gives
\begin{equation}
 \label{solat}
 \e = \Om_{\at} \e_1 \, , \qquad \Om_{\at} = \exp \left( - \frac{1}{2} \tanh^{-1} \Big( \frac{1 - \l}{1 + \l} \tanh\at \Big) \G^{01} \s_3 \right) \, .
\end{equation}
Here $\e_1$ is a spinor that depends on $(\bt , \, \gt , \, \a_1 , \, \b_1 , \, \g_1 , \, \a_2 , \, \b_2 , \, \g_2)$. Moving to the $\bt$ component \eqref{btcomponent}, the $\at$ dependence drops out if we make use of \eqref{solat} where we find
\begin{equation}
 0 = \partial_{\bt} \e_1 - \frac{1}{2} \G^{02} \s_3 \, \e_1 \, .
\end{equation}
Again, a simple integration gives
\begin{equation}
 \label{solbt}
 \e_1 = \Om_{\bt} \e_2 \, , \qquad \Om_{\bt} = \exp\Big( \frac{\bt}{2} \G^{02} \s_3 \Big) \, ,
\end{equation}
where now $\e_2$ is a spinor that depends on $(\gt , \, \a_1 , \, \b_1 , \, \g_1 , \, \a_2 , \, \b_2 , \, \g_2)$. Similarly, if we combine \eqref{solat} and \eqref{solbt}, the $\gt$ component \eqref{gtcomponent} becomes independent of $\at$ and $\bt$
\begin{equation}
 0 = \partial_{\gt} \e_2 + \frac{1}{2} \G^{12} \s_3 \, \e_2 \, .
\end{equation}
Once more, this can also be easily integrated to
\begin{equation}
 \label{solgt}
 \e_2 = \Om_{\gt} \e_3 \, , \qquad \Om_{\gt} = \exp\Big( -\frac{\gt}{2} \G^{12} \s_3 \Big) \, ,
\end{equation}
with $\e_3$ being a spinor that depends on $(\a_1 , \, \b_1 , \, \g_1 , \, \a_2 , \, \b_2 , \, \g_2)$.

The next three components \eqref{a1component}, \eqref{b1component} and \eqref{g1component} can be solved along the same lines. This process restricts the spinor $\e_3$ to the form
\begin{equation}
 \label{sola1b1g1}
 \e_3 = \Om_{\a_1} \Om_{\b_1} \Om_{\g_1} \e_4 \, ,
\end{equation}
where $\e_4$ is a spinor that depends on $(\a_2 , \, \b_2 , \, \g_2)$ and
\begin{equation}
 \begin{aligned}
  & \Om_{\a_1} = \exp \left( - \frac{1}{2} \tan^{-1} \Big( \frac{1 + \l}{1 - \l} \tan\a_1 \Big) \G^{45} \s_3 \right) \, ,
  \\[5pt]
  & \Om_{\b_1} = \exp \Big( \frac{\b_1}{2} \G^{34} \Big) \, , \quad \Om_{\g_1} = \exp \Big( \frac{\g_1}{2} \G^{45} \Big) \, .
 \end{aligned}
\end{equation}

Finally, turning to the last three gravitino components, we observe that they can be obtained from those along $(\a_1, \, \b_1 , \, \g_1)$ by mapping
\begin{equation}
 (\a_1 , \, b_1 , \, \g_1) \mapsto (\a_2 , \, \b_2 , \, \g_2) \, , \qquad (\G^3 , \, \G^4 , \, \G^5) \mapsto (\G^6 , \, \G^7 , \, \G^8) \, .
\end{equation}
This suggests that the spinor $\e_4$ takes the form
\begin{equation}
 \label{sola2b2g2}
 \e_4 = \Om_{\a_2} \Om_{\b_2} \Om_{\g_2} \eta \, ,
\end{equation}
where $\eta$ is a constant spinor and
\begin{equation}
 \begin{aligned}
  & \Om_{\a_2} = \exp \left( - \frac{1}{2} \tan^{-1} \Big( \frac{1 + \l}{1 - \l} \tan\a_2 \Big) \G^{78} \s_3 \right) \, ,
  \\[5pt]
  & \Om_{\b_2} = \exp \Big( \frac{\b_2}{2} \G^{67} \Big) \, , \quad \Om_{\g_2} = \exp \Big( \frac{\g_2}{2} \G^{78} \Big) \, .
 \end{aligned}
\end{equation}

In summary, if we combine \eqref{solat}, \eqref{solbt}, \eqref{solgt}, \eqref{sola1b1g1} and \eqref{sola2b2g2} we find that the Killing spinor $\e$ is
\begin{equation}
 \label{KillingSpinorFinal}
 \e = \Om_{\at} \Om_{\bt} \Om_{\gt} \Om_{\a_1} \Om_{\b_1} \Om_{\g_1} \Om_{\a_2} \Om_{\b_2} \Om_{\g_2} \eta \, .
\end{equation}

Before we conclude, we would like to restate the projection conditions \eqref{Projectionlambda0} and \eqref{Projectionlambdane0} from the point of view of the constant spinor $\eta$. As a consistency check, one expects to retrieve two algebraic conditions with no dependence on the coordinates. Starting with \eqref{Projectionlambda0}, we notice that both $\G^{012345}$ and $\G^{012678}$ pass freely through the $\Om$ matrices in \eqref{KillingSpinorFinal}. Therefore, equation \eqref{Projectionlambda0} simply reduces to \eqref{Projectionlambda0eta}. The case \eqref{Projectionlambdane0} is more delicate. For convenience we write the projector $\mathbb{P}_2$ as
\begin{equation}
 \mathbb{P}_2 = - \Om^2_{\at} \G^2 \Om^2_{\a_1} \G^{345} \Om^2_{\a_2} \G^{678} (i \s_2) \, .
\end{equation}
Using the commutation properties for the $\Om$ matrices it is easy to verify that the projection condition \eqref{Projectionlambdane0} implies
\begin{equation}
 \G^{2345678} (i \s_2) \eta = \eta \, .
\end{equation}
If we combine the above with the chirality condition for the type-IIA supergravity, we recover \eqref{Projectionlambdane0eta}.



\end{document}